# Light-Metal Functionalized Boron Monoxide Monolayers as Efficient Hydrogen Storage Material: Insights from DFT Simulations

Wael Othman[1,2], Wadha Al Falasi[3,4], Tanveer Hussain[5], and Nacir Tit[3,4*]

[1] Engineering Division, New York University Abu Dhabi, Abu Dhabi, United Arab Emirates

[2] Mechanical and Aerospace Engineering, Tandon School of Engineering, New York University, NY 10012, USA

[3] Physics Department, UAE University, Al Ain, United Arab Emirates

[4] Water Research Center, UAE University, Al-Ain, United Arab Emirates

[5] School of Science and Technology, University of New England, Armidale, New South Wales, 2351, Australia

*Corresponding Author: ntit@uaeu.ac.ae

## Abstract

Exceptionally high energy density by mass, natural abundance, widespread applications, and environmental friendliness make hydrogen ($H_2$) a front-runner among clean energy options. However, the transition toward clean and renewable energy applications and the actualization of $H_2$ economy require an efficient $H_2$ storage medium. Material-based $H_2$ storage is a viable option, as liquefaction and storage under pressure require ultra-low temperature (-253 ºC) and tremendously high pressure (700 atm), respectively. In this work, we highlight the exceptional $H_2$ storage capabilities of recently synthesized boron monoxide (BO) monolayer functionalized with light metals (Li, Na, K, and Ca). Our computational approach, employing density functional theory (DFT), *ab initio* molecular dynamics (AIMD), and thermodynamic analysis, reveals promising results. We found that up to four metal dopants (Li, Na, K, and Ca) can be adsorbed onto BO monolayer with significantly strong binding energies (-2.02, -1.53, -1.52, and -2.24 eV per dopant, respectively). Importantly, these bindings surpass the cohesive counterparts of the parental metal bulks, consequently stabilizing the crystal integrities, as confirmed by AIMD simulations. Each metal dopant on BO efficiently adsorbs multiple $H_2$ molecules through electrostatic and van der Waals interactions. Interestingly, the metal-functionalized BO monolayers exhibit exceptionally high $H_2$ gravimetric capacities of 11.75, 9.52, 9.80, and 11.43 wt% for 4Li, 4Na, 4K, and 4Ca@BO, respectively. These promising capacities exceed the 5.50 wt% target set by the US Department of Energy for 2025. Following the same guidelines, the average binding energy per $H_2$ molecule is within the range of -0.17 to

−0.32 eV. The adsorption and desorption of $H_2$ under practical working conditions are investigated by Langmuir adsorption model based statistical thermodynamic analysis, further supporting the potential of metal-functionalized BO monolayers for material-based $H_2$ storage applications.

**PAC Numbers:** 31.15.E; 68.43.-h; 68.43.Mn; 68.65.Pq; 73.43.Cd; 75.70.Rf

## I. Introduction

The substantial increase in global energy demand has led to a 1.0 °C rise in global temperature and a simultaneous increase in sea levels, primarily attributed to excessive fossil fuel usage and its associated greenhouse gas emissions [1]. In an effort to address global warming and climate change challenges, renewable and clean energy resources like solar, wind, and hydropower are substituting fossil fuels. These sustainable alternatives offer the dual benefits of fulfilling future energy demands and preserving the environment [2]. However, the intermittent nature of sustainable energy sources underscores the crucial need for energy storage solutions with high efficiency, sustainability, and eco-friendliness. In contrast to traditional energy storage, hydrogen ($H_2$) has emerged as the most promising candidate thanks to its ultra-high energy density by mass, abundance, and environmental friendliness [3]. The production of $H_2$ through water electrolysis only emits oxygen ($O_2$), and its combustion in fuel cells produces solely water ($H_2O$), making it a carbon-free energy carrier [4-6]. As fuel cell vehicles (FCVs), including cars, buses, trains, and airplanes, have entered the race towards sustainable transportation, their effective implementation relies heavily on practical $H_2$ storage, a key focus in current research themes [7].

The challenge of $H_2$ storage arises from its low density in gaseous form at ambient pressure and temperature, approximately 1/14th compared to air density [8]. Conventional $H_2$ storage methods typically involve pressurized tanks or liquification. However, these technologies raise safety concerns and struggle to store hydrogen cost-effectively for extended periods. For instance, storing hydrogen as a gas typically requires gigantic tanks with high pressures ranging from 350 to 700 bar to hold a practical amount of $H_2$ [8]. Meanwhile, FCVs currently rely on pressurized $H_2$ in vessels, requiring an average of 9 kg of $H_2$ for a 500 km driving range [8]. The intrinsic limitations of pressurized tank storage result in insufficient volumetric and gravimetric storage densities, along with safety considerations. On the other hand, storing hydrogen as a liquid requires cooling it down to cryogenic temperatures (−252.8 °C).

Concerns regarding safety and overall cost in conventional $H_2$ storage have urged the exploration of alternative storage technologies [9-10]. In this context, material-based (solid-state) $H_2$ storage becomes the most viable option to minimize the volume and power required for practical $H_2$ storage conditions. Ideally, storage materials should achieve high storage capacity, rapid chemical kinetics, and a reversible low dehydrogenation temperature. These characteristics depend on the number of chemically reactive sites and binding between the host material and $H_2$ molecules. In this regard, materials with a high surface area are advantageous in maximizing the number of stored $H_2$ molecules. At the same time, the type of binding (physisorption or chemisorption) influences the chemical kinetics and adsorption-desorption reversibility. According to the Langmuir equation, moderate binding energies of $H_2$ molecules on the storage medium should be around 0.155 eV/atom, or 14.96 kJ/mol, to achieve appropriate reversibility under practical conditions around 30 bar and 298 K [11]. Additionally, a moderate interaction between the host material and $H_2$ molecules is required to ensure strong physisorption, with adsorption energy in the 0.15–0.60 eV/$H_2$ range, along with sufficient gravimetric density to surpass the United States Department of Energy (US-DoE) goal of 5.5 wt% [12, 13].

Various materials, including metal hydrides and porous nanomaterials, have been proposed as promising mediums for $H_2$ storage, where hydrogen is stored either on the surfaces (adsorption) or within solids (absorption). However, inherent limitations, such as low storage capacity, slow chemical kinetics, and elevated reversible dehydrogenation temperatures, hinder their practical applications [13]. Among the available options, two-dimensional (2D) materials stand out due to their lower mass density and ultra-high surface areas reaching 2078 $m^2$/g for boron nitride (BN) [14] and 2630 $m^2$/g for graphene [15]. While the pristine forms of the most 2D materials weakly interact with $H_2$ molecules [16-19], doping with alkali, alkaline-earth, or transition metals serves as a promising approach to intensify $H_2$ bindings [20-23]. The low electronegativities of such metals induce strong polarization, enhancing the attraction of $H_2$ molecules (electrostatic and van der Waals interactions) and overall $H_2$ storage performance, as demonstrated with functionalized $B_2O$ [24], graphene [25], haeckelite [26], Ene–Yne [27], $C_2N$ [28, 29], BN [30], and phosphorene [31].

Recent literature reports significant progress in utilizing 2D materials for $H_2$ storage. For instance, polar 2D materials, particularly functionalized MXenes and transition metal dichalcogenides (TMDs), appear to be highly promising [10, 32-34]. Kumar and co-workers [10] have shown that $Ti_2C$ MXene can achieve an uptake capacity of up to 8.6 wt%, which is higher than the gravimetric capacity of metal-based complex hybrids (~5.5 wt%). Zhu et al. [35] demonstrated the high potential of TMDs as cost-effective catalysts for the hydrogen evolution reaction (HER), showcasing stable adsorption of a substantial number of $H_2$ molecules on 1T-$MoS_2$ with appropriate adsorption energy. In a separate study, Wang *et al.* [36] discovered that $MoS_2$ nanotubes (NT) effectively attract $H_2$ and methane molecules to both outer and inner surfaces, reporting hydrogen storage capacity in the range of 0.7-0.9 wt% depending on the nanotube diameter. Recently, Alfalasi *et al.* showed that doping the Se sites in the combined $MoSe_2$/Graphene bilayer with Mn improves the gravimetric $H_2$ capacity, primarily attributed to enhanced electric dipole moments in the TMD monolayer [32]. These studies underscore the growing interest in utilizing 2D materials for $H_2$ storage applications, highlighting their potential to propel advancements in energy storage technologies.

Nevertheless, 2D materials consisting of heavy elements exhibit limited $H_2$ storage capacities, such as GeC [37], SnC [38], and $MoS_2$ [39]. On the other hand, certain carbon-based nanostructures, such as carbon nanotubes [40], fullerene [40, 41], and graphene [42]. show weak interactions towards $H_2$ molecules, and result in low storage capacities. Heavy transition metals usually possess high cohesive energies and favor clustering instead of homogenously distributing over the storage materials, resulting in sluggish reversibility and low storage capacity [43]. Considering these factors, 2D materials composed of light atoms and functionalized with light alkali or alkaline-earth metals emerge as ideally feasible candidates for achieving high $H_2$ storage capacities.

In 2023, Perras *et al.* [44] employed advanced multidimensional nuclear magnetic resonance methods to unravel the atomic structure of boron monoxide (BO), a material initially reported in 1940 [45] and synthesized in 1955 [46]. The recent material characterization revealed the presence of interconnected $B_4O_2$ rings forming two-dimensional nanosheets of BO [44]. The large pores within BO act as attraction sites for alkali and alkaline-earth metals. This makes the material highly promising for $H_2$ storage, representing a potential application that had not been previously explored. Motivated by these intriguing properties, this study showcases the significant potential of recently synthesized BO monolayers in $H_2$ storage applications using the Vienna Ab initio Simulation Package (VASP). Our theoretical analysis utilizes state-of-the-art tools, including density-functional theory (DFT), *ab initio* molecular dynamics (AIMD), and thermodynamic analysis, to assess effective hydrogenation uptake capacity. By embedding multiple light metal atoms (Li, Na, K, and Ca) into the large pores of BO sheets, our findings indicate that the ionic bindings exceed the cohesive energies of the corresponding bulk metals. This prevents metal clustering and stabilizes the crystal integrity of metal-functionalized BO, making it well-suited for $H_2$ storage with high gravimetric uptake capacities of up to 11.75 wt%. As a result, metal-functionalized BO emerges as a promising class of non-precious 2D materials for hydrogen storage functionality, filling the gap in materials-based clean energy storage.

## II. Computational Methods

To unravel the H$_2$ storage capabilities of BO, we conducted spin-polarized DFT calculations involving atomic structure optimization, adsorption mechanism, charge transfer analysis, and electronic properties assessments. These calculations were performed using the VASP within the frameworks of DFT and the projected augmented plane wave (PAW) method [47]. The Perdew-Burke-Ernzerhof form of the generalized gradient approximation (PBE-GGA) approach was used to address the exchange-correlation energy [48]. Given that weakly-dispersive van der Waals interactions play a crucial role in the intrinsic attractive forces, we incorporated the DFT-D3 method of Grimme to obtain reliable adsorption energies [49]. The modeled structures throughout this study were based on the relaxed unit cell of BO, having square 2D-Bravais lattice structure with lattice parameter $a = b$ = 7.826 Å and a basis of eight B and eight O atoms. Adequate 15-Å vacuum space was added to the vertical direction of the BO to eliminate the self-interactions among the periodically adjacent layers. For the structural relaxation, we applied the convergence tolerances for the total energy of 10$^{-6}$ eV and atomic force of 0.01 eV/ Å. In sampling the Brillouin zone, we utilized the Monkhorst-Pack technique with 5×5×1 K-grid during geometry optimization [50]. A denser K-point grid of 8x8x1 was used for the density of states (DOS) calculations. The cut-off energy of the plane-wave was 500 eV. Bader-charge analysis within the framework of VASP was applied to calculate the charge exchange between H$_2$ molecules and the host medium [51].

The average binding energy ($E_{bind}$) of a metal atom ($M$) embedded in the pores of BO monolayer substrate were calculated using the formula:

$$E_{bind} = \frac{E_{BO+nM} - (E_{BO} + nE_M)}{n} \quad (1),$$

where $E_{BO+nM}, E_{BO}, E_M$ stand for the total energies of the system of BO embedded with $n$ metal atom(s), pristine BO, and isolated metal atom, respectively. Moreover, *ab initio* molecular dynamics (AIMD) was performed to further validate the structural stabilities of metal-functionalized BO at 400 K. Nose-Hoover thermostat was employed to control the temperature, with a time step of 1.0 fs and for 8 ps.

The average adsorption energy ($E_{ads}$) of H$_2$ molecule on top of the substrate is defined as:

$$E_{ads} = \frac{E_{Sheet+mH_2} - (E_{Sheet} + mE_{H_2})}{m} \qquad (2),$$

where, $E_{Sheet+mH_2}$, $E_{Sheet}$, $E_{H_2}$ are the total energies of the $m$H₂ molecule(s) adsorbed on the BO (adsorbent), the BO, and the isolated H₂ molecule, respectively. The uptake gravimetric capacity is defined as follows [32]:

$$C_T(wt\%) = \left[\frac{N_T \cdot M(H_2)}{N_T \cdot M(H_2) + N_{metal} \cdot M(metal) + N_{BO} \cdot M(BO)}\right] \times 100\% \qquad (3),$$

where $M(H_2)$, $M(\text{metal})$, and $M(BO)$ stand for the molecular masses of H₂ molecule, metal adatom, and host crystal BO, whereas $N_T$, $N_{metal}$, and $N_{BO}$ stand for the number of H₂ molecules, metal adatoms, and BO pairs in the substrate.

The adsorption-desorption characteristics at operational conditions were analyzed by statistical thermodynamics according to the grand canonical partition function ($z$):

$$z = 1 + \sum_{i=1}^{n} e^{-\frac{(E_b^i - \mu)}{k_B T}} \qquad (4),$$

where $n$ represents the maximum number of adsorbed H₂ molecules, whereas $E_b^i$, $k_B$, $\mu$, and $T$ represent the adsorption energy of the $i^{th}$ adsorbed H₂ molecule, the Boltzmann constant (1.38 x 10⁻²³ J/K), chemical potential of the gas phase of the H₂ molecule, and absolute temperature, respectively. In particular, $\mu$ is a function of $P$ and $T$, defined as:

$$\mu_{H_2}(P,T) = \Delta H + T\Delta S + k_B T \ln \frac{P}{P_0} \qquad (5),$$

where $\Delta H$, $\Delta S$, $P$, and $P_0$ represent the enthalpy change, entropy change, pressure, and the atmospheric pressure (1.01 x 10⁵ Pa), respectively. The values of $\Delta H + T\Delta S$ are obtained from the experimental database [43]. Meanwhile, the number of stored H₂ molecules ($N$) in the host material can be expressed by:

$$N = N_0 \left[\frac{Z-1}{Z}\right] \qquad (6),$$

where $N_0$ represents the maximum number of H₂ molecules adsorbed on the host medium at 0 K ($N_T$).

## III. Results and Discussion

### (a) Pristine BO Monolayer Properties

Pristine boron monoxide (BO) monolayer crystallizes in a square Bravais lattice with a basis comprising eight B and eight O atoms. This structure belongs to the space group P4/mmm (No. 123) [52]. **Figure 1(a)** shows the relaxed atomic structures using PBE-DFT method. The bond length B-O within 1.37-1.39 Å and the lattice constant of a = 7.83 Å are consistent with both original synthesized structure [44] and *ab initio* predictions [52]. **Figures 1(b)** and **1(c)** show the corresponding spin-polarized band structure and projected density of states (PDOS), respectively. The band structure shows that BO is wide-bandgap semiconductor with an indirect bandgap of energy $E_g$ = 2.23 eV (i.e., the valence-band maximum (VBM) is at Γ point and the conduction band minimum (CBM) is at M point of the Brillouin zone). This value of bandgap energy is underestimated as being calculated using PBE-DFT method. Yet this value is consistent with the first-principles results of 2.18 eV reported by Mortazavi et al. [52]. These latter authors further involved HSE06 hybrid functional to improve the bandgap and reported $E_g$= 3.78 eV, in agreement with the experimental value. Furthermore, the spin-polarized PDOS, shown in **Figure 1(c)**, confirms that BO is a paramagnetic wide-bandgap semiconductor. The near VBM states have equal contributions from $p_x$ and $p_x$ orbitals of both B and O atoms, whereas the near CBM states receive contributions form $p_z$ states of boron more than oxygen atoms. Hence, the PDOS reveal that BO is a covalent material with partial ionic character, as oxygen possesses more electronegativity than boron (i.e., χ$^B$ = 2.04 Pauling and χ$^O$ = 3.44 Pauling [53]).

It is customary, first, to assess the adsorption of a single H₂ molecule on pristine material. Hence, on the primitive cell shown in dashed line in Figure 1a, we carried out the relaxation of a single H₂ molecule on various possible sites, as indicated in **Figure S1(a)** (supporting information). All the adsorption processes led to physisorption types, with adsorption energies shown in chart diagram of **Figure S1(b)** (supporting information).  It seems that two adsorption sites are favorable and do correspond to the centres of the two biggest pores (X6, and X7), having adsorption energies of values $E_{ads}$ = -0.158 eV and -0.159 eV, respectively. We decided to set the threshold energy at -0.16 eV in our study of uptake capacity shown below. Of course, the threshold energy is arbitrary chosen in many research groups. Nonetheless, in our case, the value has significance; after the functionalization of BO, all absolute values of average adsorption energies larger than this threshold should be considered an improvement over the pristine case of adsorption. Meanwhile, these adsorption energies of a single H₂ molecule on pristine BO are below the energy range [-0.6, -0.2] eV set by the US-DoE as recommended for hydrogen storage applications. So, one way to enhance the adsorption energy is to consider the functionalization of BO with alkali and rare earth alkaline metals (such as Li, Na, K, and Ca). We emphasize that this selection of metal is intended to exclude the transition metals in order to avoid the chemisorption processes. So, the selected light atoms are expected to alter the bonding with the lattice and contribute to the formation of electric dipole moments to make the substrate even more polar. The enhancement in the polarity of the surface could induce dipole moments into the $H_2$ molecules, as will be indicated by the enhancement in the adsorption energy described below.

**(b) Light Metal Functionalization**

The primitive cell of BO possesses two hexagons ($B_4O_2$), one intermediate pore ($B_8O_4$) and one large pore ($B_8O_8$), as shown in **Figure 1(a)**. The light metal atoms (X = Li, Na, K, and Ca) can be embedded in the latter two pores and decorated on the former two hexagons. Hence, **Figure 2** and **Figure S2** (supporting information) show the optimized atomic structures of metal-functionalized BO. We have functionalized the BO primitive cell one by one of light metal atoms up to 4 atoms at maximum (referred as 4X@BO). As expected, the first two light metal atoms got fully embedded in the two large pores and stabilize within the same membrane as BO monolayer. Meanwhile, the next two light metal atoms were just decorated on the top of the two $B_4O_2$ hexagons (see **Figure S2** and **Figure 2**). The average $E_{bind}$ of these metal adatoms were calculated using equation (1), and the absolute values were found to be stronger than the cohesive energy of the corresponding bulk metals (i.e., $|E_{bind}^{ave}| > |E_{coh}|$, for up to 4 light metal atoms), as shown in **Figure 3(a)**. In each case, a maximum of four light metal atoms of Li, Na, K, and Ca are found bonded with the BO, resulting in strong $E_{bind}$ values of -2.02, -1.53, -1.52, and -2.24 eV per dopant, respectively. It is worth emphasizing that the average $E_{bind}$ is more relevant than the recursive $E_{bind}$ as it better simulates the experimental synthesis of functionalization more realistically. The fact that $|E_{bind}| > |E_{coh}|$ should reveal that the metal dopants are more inclined to bind to the lattice rather than clustering. This fosters the stability of light-metal functionalized BO sheets, thereby increasing their viability for hydrogen storage. While metal functionalization induces buckling in the atomic structures of BO sheets, our AIMD analysis, conducted at 400 K for 8 ps by a step of 1 fs, indicates insignificant fluctuations in the total energies during the simulation, as depicted in **Figure 3(b)**. This further confirms the structural integrity of metal-functionalized BO over time. It is noteworthy that Mg exhibited weak $E_{bind}$ to the BO sheet, leading to its exclusion from the list of selection of catalyst candidates.

Moreover, the results of the spin-polarized band structure and PDOS calculations, shown in **Figure 4**, reveal significant alterations in the electronic properties of BO upon metal functionalization, which characterize the chemical bonds and support the strong $E_{bind}$. The absence of the band gap indicates the induced metallization of all 4X@BO systems. However, in the case of 4Na and 4K@BO, split bands (half-metallicity) persist for spin-down electrons, with energy values of 0.182 and 0.230 eV, respectively.

Meanwhile, the spin-polarization calculations provide information about the magnetization, which contributes to the interaction with H$_2$ molecules, particularly in the cases of 4K, 4Li, and 4Na@BO with total magnetization of 2.109, 1.062, 0.03 $\mu_B$, respectively. **Table 1** summarizes the electronic properties (Bader charge transfer, band-gap energy, and total magnetization) of pristine BO and light-metal functionalized BO (i.e., 4X@BO, with X = Li, Na, K, and Ca).

Additionally, we conducted calculations of charge density difference (CDD) to understand the binding nature of metal atoms to the BO lattice. **Figure 5** presents the calculated CDD, showing the bond characteristics of maximally localized Wannier functions (MLWFs). In particular, the Wannier orbitals in Li- and Ca-decorated BO in **Figures 5(a)** and **5(d)** exhibit an enhanced population of the positive value (yellow) and gain of charge, whereas those in Na- and K-decorated BO in **Figures 5(b)** and **5(c)** show a more spherical shape of the negative value (cyan) and charge deficit.

**(c) Hydrogenation of 4X@BO**

The H$_2$ storage capacities of the 4X@BO are subsequently evaluated. In **Figure 6**, the optimized atomic structures showcase the maximum number of H$_2$ molecules adsorbed on the 4X@BO systems. Additionally, **Figure 7(a)** shows the absolute values of the corresponding average adsorption energies. It is worth mentioning here that H$_2$ adsorption is carried out in stepwise manner. Initially, a single H$_2$ molecule is introduced to each metal dopant, a total of 4H$_2$ on 4X@BO, and the systems are completely optimized. Next, the number of H$_2$ are increased to 2, 3, 4, and 5 per metal dopant on 4X@BO, and structural relaxations are performed in each situation. In each case, the $E_{ads}$ values (per H$_2$) are calculated by equation (2) and compared with the threshold value of -0.160 eV/H$_2$. Our results suggest that Li-, Na-, K-, and Ca-decorated BO can adsorb up to 16, 16, 20, and 24 H$_2$ molecules, respectively. These H$_2$ molecules are physisorbed on the 4X@BO *via* the predominant electrostatic and van der Waals-type interactions. The average $E_{ads}$ values are -0.205, -0.175, -0.185, and -0.204 eV for fully hydrogenated 4Li, 4Na, 4K, and 4Ca@BO, respectively, exceeding the -0.160 eV threshold in magnitude.

Subsequently, the maximum numbers of stored H$_2$ molecules (N$_T$) are used to calculate the theoretical H$_2$ gravimetric capacities (C$_T$) according to equation (3). All atomic weights are sourced from the database [54]. As shown in **Figure 7(b)**, the theoretical gravimetric capacities are 11.75, 9.52, 9.80, and 11.43 wt%, corresponding to 4Li, 4Na, 4K, and 4Ca@BO, respectively. It is important to acknowledge that DFT utilizes the Born-Oppenheimer frozen lattice approximation, wherein all calculations are performed at 0 K. These theoretical capacities exclude the essential thermodynamic contributions of finite pressure and temperature.

**(d) Thermodynamic analysis**

Since it is crucial to understand the effective H$_2$ storage capacity of the proposed materials at ambient conditions of practical temperature and pressure, we conducted a statistical thermodynamic analysis based on the Langmuir adsorption model [55-57]. **Figure 8** shows, based on equations (4-6), the number of H$_2$ molecules (N) adsorbed at given P and T values (*i.e.*, N-P-T diagram) on the BO primitive cell with 4 metal dopants (i.e., 4X@BO, X = Li, Na, K, and Ca). Remarkably, the H$_2$ molecules can be released from the storage material under high-temperature and low-pressure conditions, or else they can be stored in low-temperature and high-pressure conditions. Furthermore, one can assess the realistic absorption-desorption characteristics of the adsorbed H$_2$ molecules. In practice, the pressure (P) and temperature (T) of adsorption are 30 atm and 25 ºC, respectively whereas those of desorption are 3 atm and 100 ºC, respectively [58]. Accordingly, **Table 2** compiles the calculated theoretical storage at 0 K ($C_T$) and effective storage ($C_E$) at the practical adsorption and desorption conditions. The obtained effective H$_2$ gravimetric capacities of 4Li, 4Na, 4K, and 4Ca@BO are 9.88, 3.96, 8.53, and 9.78 wt%, respectively. Notably, the capacities of BO sheets decorated with 4Li, 4K, and 4Ca surpass the goal value of 5.5 wt% set by the US-DoE for 2025 [24]. In **Table 3**, we present a comprehensive comparison between our results using DFT and those from similar DFT studies of different 2D materials reported in the literature. Overall, our theoretical gravimetric capacity of 4X@BO is almost always higher than in many DFT-studied materials. Besides, it is remarkable that in the few cases of materials whose gravimetric capacities are higher than ours, the authors carried out the hydrogenation to much lower threshold average adsorption energies. Thus, our 4X@BO systems can potentially surpass these systems by accommodating more H$_2$ molecules and lowering the average $E_{ads}$ even further below our set threshold energy (-0.16 eV/ H$_2$ molecule). Based on our findings, it is thus conclusive that BO sheets functionalized with alkali and alkaline-earth metals are promising materials for hydrogen storage applications.

## IV. Conclusions

In this study, we demonstrated the remarkable H$_2$ storage efficiency of BO monolayers functionalized with Li, Na, K, Ca. Utilizing a comprehensive set of theoretical approaches, including DFT, AIMD, and thermodynamic analysis, we investigated the theoretical and effective H$_2$ storage capacities of 4X@BO at ambient conditions. Our findings revealed that the presence of pores in BO monolayer induces robust ionic bonds with all the light metal dopants, while having average binding energies that surpass the cohesive energies of the corresponding bulk metals. This criterion simultaneously prevents the formation of metal clusters and guarantees the stable crystal integrity of 4X@BO. The thermal stabilities of 4X@BO were confirmed through AIMD at 400K. The semiconducting to metallic transition of BO upon the metal functionalization was caused by the significant charge transfer from the dopants to the monolayer. We found that 4X@BO adsorbed a large number of $H_2$ with desirable average adsorption energies of -0.205, -0.175, -0.185, and -0.204 eV for fully hydrogenated 4Li, 4Na, 4K, and 4Ca@BO, respectively. It was observed that Li- and Ca-decorated BO monolayer exhibit theoretical $H_2$ storage capacities of 11.75 and 11.43 wt%, respectively, exceeding the US-DoE goal of 5.50 wt% by 2025. Consequently, 4X@BO emerges as a promising class of 2D materials for $H_2$ storage applications. The insights gained from our theoretical calculations will guide the synthesis of efficient materials for high reversible capacity in $H_2$ storage applications at practical operating conditions.

## Acknowledgements

The authors are indebted to thank Dr. Thomas Fowler for critical reading of the manuscript and the National Water and Energy Center (NWEC) at the UAE University for the financial support (Grant number: 12R162).

## Author Contributions:

**Wael Othman:** Investigation, Formal analysis, Methodology, Visualization, Validation, Project administration, Software, Writing - original draft preparation. **Wadha Al-Falasi:** Investigation, Formal analysis, Methodology, Validation. **Tanveer Hussain:** Conceptualization, Investigation, Formal analysis, Methodology, Resources, Writing- Reviewing and Editing. **Nacir Tit:** Conceptualization, Funding acquisition, Resources, Supervision, Writing- Reviewing and Editing.

## Competing Interests:

The authors declare that they have no known competing financial interests or personal relationships that could have appeared to influence the work reported in this paper.

## Supplementary Documents:

**Figure S1: (a)** Relaxed atomic structures (top and side views) and **(b)** the corresponding adsorption energy of one $H_2$ molecule on different sites of pristine BO.

**Figure S2:** Optimized atomic structures (top and side views) of BO decorated with 1, 2, and 3 ad-atoms of **(a)** Li, **(b)** Na, **(c)** K, and **(d)** Ca (primitive cell represented by the dashed box).

## Figure Caption

**Figure 1: (a)** Optimized atomic structure (top and side views) of pristine Boron-Monoxide (Pr-BO) monolayer. The dashed box represents the primitive cell as well as the simulation supercell, which belongs to a square lattice with lattice parameter a = 7.826 Å and a basis of eight B + eight O atoms. **(b)** Band structure of Pr-BO showing an indirect band gap character of $E_g$= 2.23 eV (X = 0.401 Å$^{-1}$, M = 0.803 Å$^{-1}$, and Γ = 1.371 Å$^{-1}$). **(c)** Spin-polarized projected density of states of Pr-BO showing a paramagnetic semiconductor. The Fermi level has been shifted to zero and denoted by the horizontal dashed line.

**Figure 2:** Optimized atomic structures (top and side views) of four-metal dopants per primitive cell of BO monolayer: **(a)** 4Li@BO, **(b)** 4Na@BO, **(c)** 4K@BO, and **(d)** 4Ca@BO. The dashed box represents the primitive cell as well as the simulation supercell.

**Figure 3: (a)** Average binding energy per dopant atom of different metals compared to the cohesive energy and **(b)** Ab-initio Molecular Dynamics (AIMD) simulation plots of BO decorated with 4Li, 4Na, 4K, and 4Ca. These AIMD simulations were performed at 400 K.

**Figure 4:** Spin-polarized band structure and projected density of states (PDOS) of BO decorated with **(a)** 4Li, **(b)** 4Na, **(c)** 4K, and **(d)** 4Ca, where the X, M, and Γ k-vectors are 0.401, 0.803, and 1.371 Å$^{-1}$, respectively. The shown electronic structures confirm the metallization of BO monolayer by metal embedment. The Fermi level has been shifted to zero and denoted by the horizontal dashed line. Bands of spin-up states (solid orange lines) and of spin-down states (dotted blue lines).

**Figure 5:** Charge Density Difference (CDD) showing the bond characteristics of maximally localized Wannier functions (MLWFs) of BO decorated with **(a)** 4Li, **(b)** 4Na, **(c)** 4K, and **(d)** 4Ca, where yellow and cyan isosurfaces indicate charge aggregation and depletion, respectively. Here, the isosurface level is 0.005 e/Bohr$^3$.

**Figure 6:** Optimized atomic structures (top and side views) of the maximum hydrogen adsorption on the metal-decorated BO: **(a)** 16 $H_2$ on 4Li@BO, **(b)** 16 $H_2$ on 4Na@BO, **(c)** 20 $H_2$ on 4K@BO, and **(d)** 24 $H_2$ on 4Ca@BO.

**Figure 7: (a)** Average adsorption energy per $H_2$ molecule adsorbed on the metal-decorated BO and **(b)** the corresponding gravimetric density of adsorbed $H_2$ (wt%).

**Figure 8:** Thermodynamics analysis showing the average number of $H_2$ molecules ($N_{ave}$) adsorbed on **(a)** 4Li@BO, **(b)** 4Na@BO, **(c)** 4K@BO, and **(d)** 4Ca@BO as a function of applied temperature and pressure.

## Table Caption

**Table 1:** Electronic properties of pristine and metal-functionalized BO. $\Delta q$, $E_g$, and M represent the Bader charge transfer, band gap, and total magnetization, respectively.

**Table 2:** The optimal theoretical numbers of $H_2$ molecules ($N_T$) adsorbed on the host storage materials: 4Li@BO, 4Na@BO, 4K@BO, and 4Ca@BO. These numbers are obtained from the DFT calculations and are used to compute the theoretical storage ($C_T$) capacities. Meanwhile, $N_a$ and $N_d$ represent the amount of adsorbed $H_2$ molecules under the adsorption (P = 30 atm and T = 25 ºC) and desorption (3 atm and T =

100 ºC) conditions, respectively (equations 4-6). The practical number of $H_2$ molecules ($N_p$) that can be reversibly stored/released is given by $N_a$-$N_d$. The effective storage capacity ($C_E$) is then computed based on $N_p$ using equation (3).

**Table 3:** Comparison between different hydrogen storage systems in terms of theoretical number of adsorbed $H_2$ molecules per simulation cell ($N_T$), average adsorption energy per $H_2$ molecule ($E_{ads}$), and theoretical $H_2$ gravimetric capacity ($C_T$).

## Table 1

| System | $\Delta q$ (e/MI) | $E_g$ (eV) | M ($\mu_B$) |
|---|---|---|---|
| Pristine BO | NA | 2.183 (↑)(↓) | 0.00 |
| BO:4Li | 0.695 | Metallic | 1.062 |
| BO:4Na | 0.479 | 0.000 (↑) <br> 0.182 (↓) | 0.030 |
| BO:4K | 0.462 | 0.000 (↑) <br> 0.230 (↓) | 2.109 |
| BO:4Ca | 0.856 | Metallic | 0.00 |

## Table 2

| Storage material | $C_T$ (wt%) | $N_T$ (molecule) | $N_a$ (molecule) | $N_d$ (molecule) | $N_p$ (molecule) | $C_E$ (wt%) |
|---|---|---|---|---|---|---|
| 4Li@BO | 11.75 | 16 | 15.53 | 2.07 | 13.46 | 9.88 |
| 4Na@BO | 9.52 | 16 | 6.77 | 0.13 | 6.65 | 3.96 |
| 4K@BO | 9.80 | 20 | 18.16 | 1.10 | 17.06 | 8.53 |
| 4Ca@BO | 11.43 | 24 | 23.37 | 3.68 | 19.75 | 9.78 |

**Table 3**

| Hydrogen Storage System | n | $E_{ads}$ (eV) | $C_T$ (wt%) |
|---|---|---|---|
| Li-decorated BO (this work) | 16 | -0.205 | 11.75 |
| Na-decorated BO (this work) | 16 | -0.175 | 9.52 |
| K-decorated BO (this work) | 20 | -0.185 | 9.80 |
| Ca-decorated BO (this work) | 24 | -0.204 | 11.43 |
| V-decorated 2DPA-I [59] | 7 | -0.44 | 7.29 |
| V-decorated biphenylene [60] | 7 | -0.47 | 10.30 |
| Ti-decorated graphene [61] | 8 | -0.42 | 7.80 |
| Li-decorated B@$r_{57}$haeckelite [26] | 12 | -0.16 | 10.00 |
| Sc-doped Holey graphyne [62] | 5 | -0.36 | 9.80 |
| Li-decorated B-doped siligene [63] | 4 | -0.17 | 12.71 |
| Ti-decorated $C_2N$ [29] | 10 | -0.28 | 6.80 |
| Zr-decorated biphenylene [64] | 9 | -0.40 | 9.95 |
| Li-decorated biphenylene [65] | 12 | -0.20 | 7.40 |
| Li-decorated $B_2S$ honeycomb [66] | 12 | -0.14 | 9.10 |
| Y-decorated $C_{48}B_{12}$ [67] | 72 | -0.46 | 7.51 |
| Li-decorated MOF-5 [68] | 18 | _ | 4.30 |
| Y-decorated covalent triazine frameworks [69] | 7 | -0.33 | 7.30 |
| Sc-decorated g-$C_3N_4$ [70] | 7 | -0.39 | 8.55 |
| Ti-decorated graphene [71] | 8 | -0.46 | 6.11 |
| Y-decorated g-$C_3N_4$ [72] | 9 | -0.33 | 8.55 |
| Ti-decorated B-doped twin-graphene [73] | 8 | -0.20 | 4.95 |
| Co-decorated N-doped graphene [74] | 28 | -0.19 | 11.36 |
| V-decorated porous graphene [75] | 6 | -0.56 | 4.58 |
| Ti-decorated graphene [76] | 8 | -0.21 | 6.30 |
| K-doped $PC_{71}BM$ [77] | 45 | -0.14 | 6.22 |
| Ca-decorated DCV graphene [78] | 14 | -0.10 | 5.80 |
| Li-decorated graphene nanoribbons [79] | 8 | -0.24 | 3.80 |
| Li-decorated $T_{4,4,4}$-graphyne [80] | 16 | -0.20 | 10.46 |
| K-decorated Ga-doped germanene [81] | 36 | -0.19 | 8.19 |
| Li-decorated P-$BN_2$ [82] | 28 | -0.16 | 13.27 |
| Li-decorated N-doped penta-graphene [83] | 12 | -0.24 | 7.88 |
| Li-decorated defective penta-$BN_2$ [84] | 16 | −0.14 | 9.17 |
| Li-decorated $B_3S$ [85] | 12 | −0.17 | 7.70 |
| Li-decorated penta-silicene [86] | 12 | -0.22 | 6.42 |
| Li-decorated penta-octa-graphene [87] | 3 | -0.22 | 9.90 |
| K-decorated SnC [38] | 6 | -0.20 | 5.50 |
| Li-decorated borophene [88] | 10 | -0.36 | 9.00 |
| Li-decorated boron phosphide [89] | 16 | -0.19 | 7.40 |
| Li-decorated defected biphenylene [7] | 14 | -0.20 | 8.75 |

# Figures

# Light-Metal Functionalized Boron Monoxide Monolayers as Efficient Hydrogen Storage Material: Insights from DFT Simulations

Wael Othman[1,2], Wadha Al Falasi[3,4], Tanveer Hussain[5], and Nacir Tit[3,4*]

[1]Engineering Division, New York University Abu Dhabi, Abu Dhabi, United Arab Emirates

[2]Mechanical and Aerospace Engineering, Tandon School of Engineering, New York University, NY 10012, USA

[3]Physics Department, UAE University, Al Ain, United Arab Emirates

[4]Water Research Center, UAE University, Al-Ain, United Arab Emirates

[5]School of Science and Technology, University of New England, Armidale, New South Wales, 2351, Australia

*Corresponding Author: ntit@uaeu.ac.ae

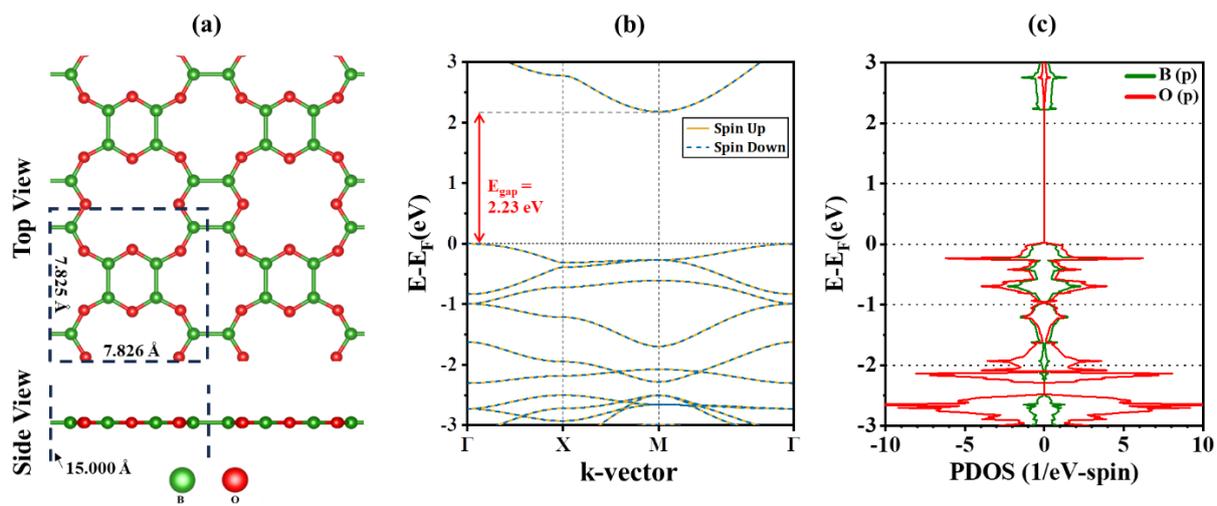

**Figure 1**

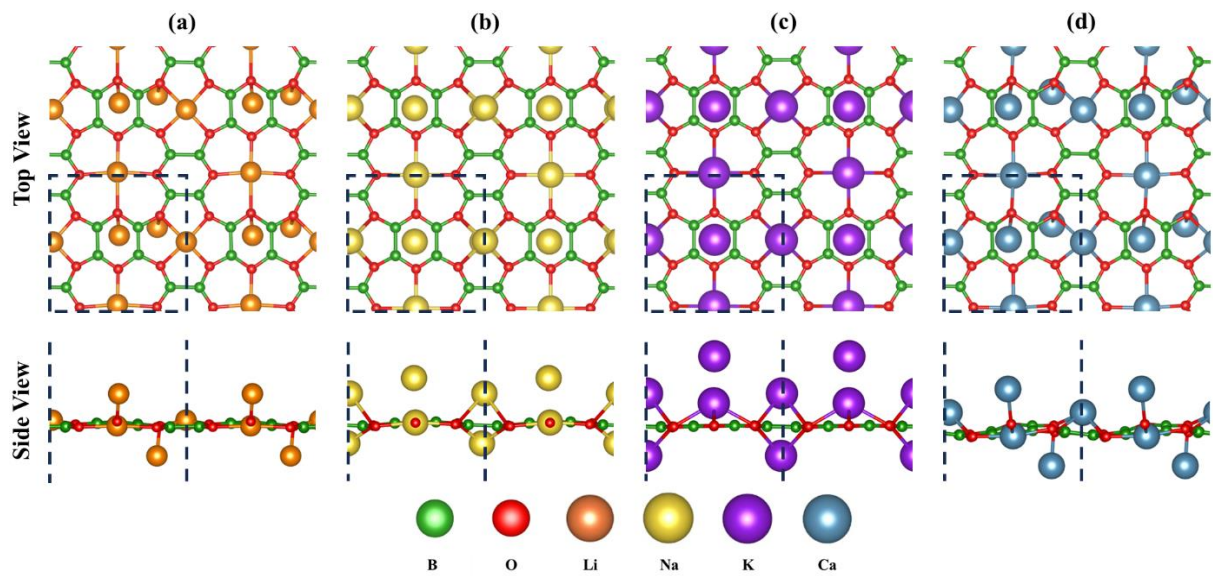

**Figure 2**

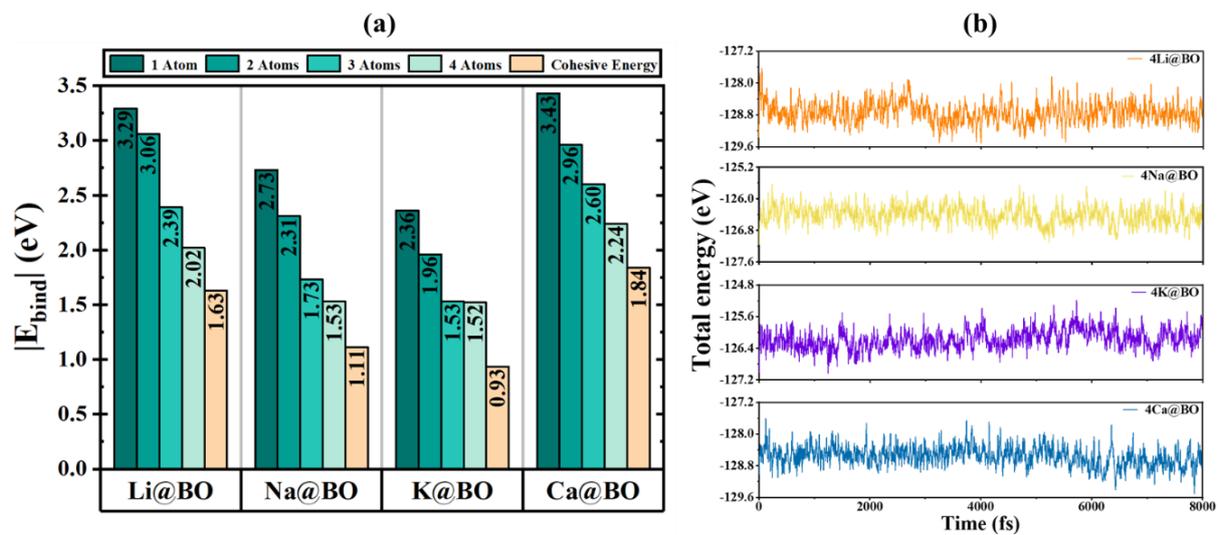

**Figure 3**

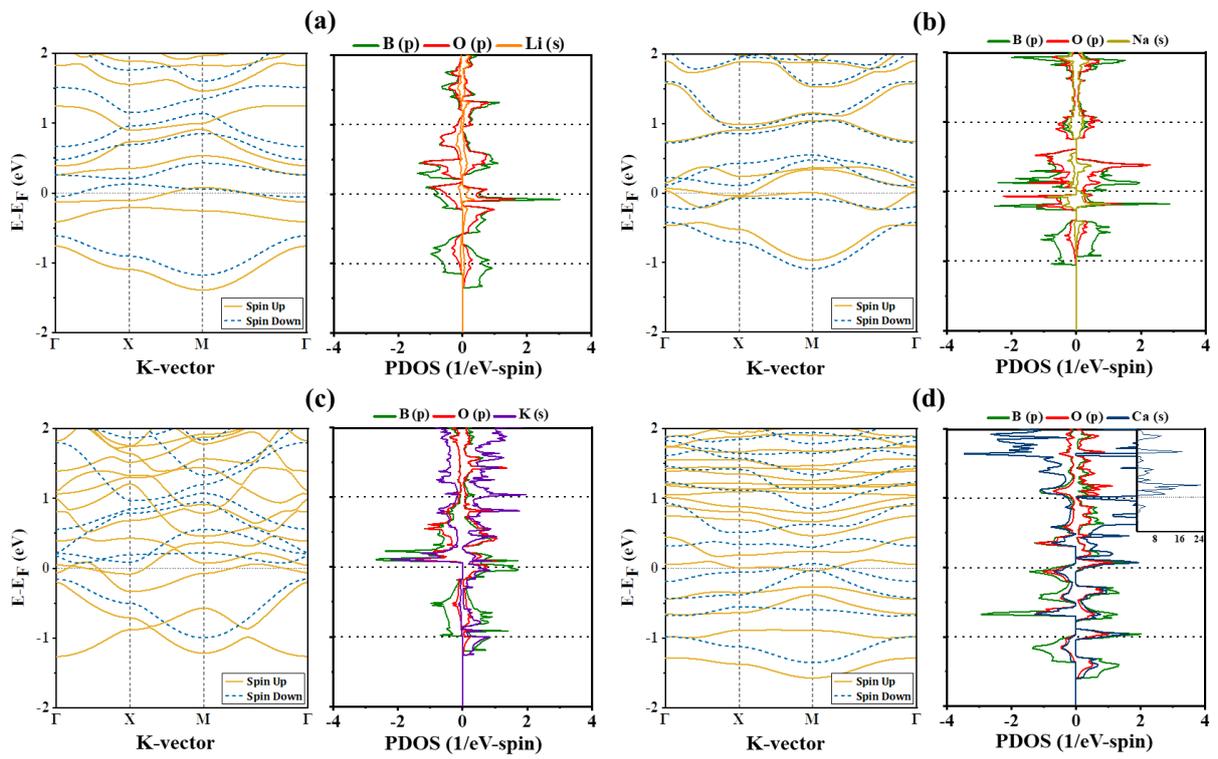

**Figure 4**

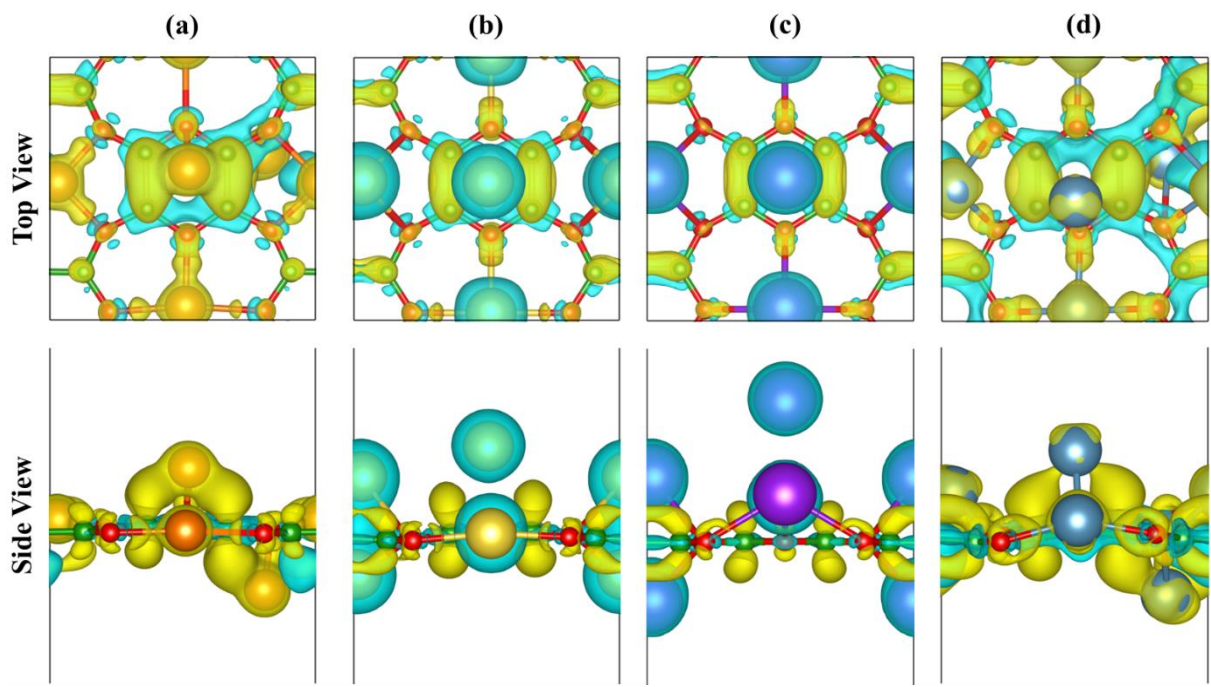

**Figure 5**

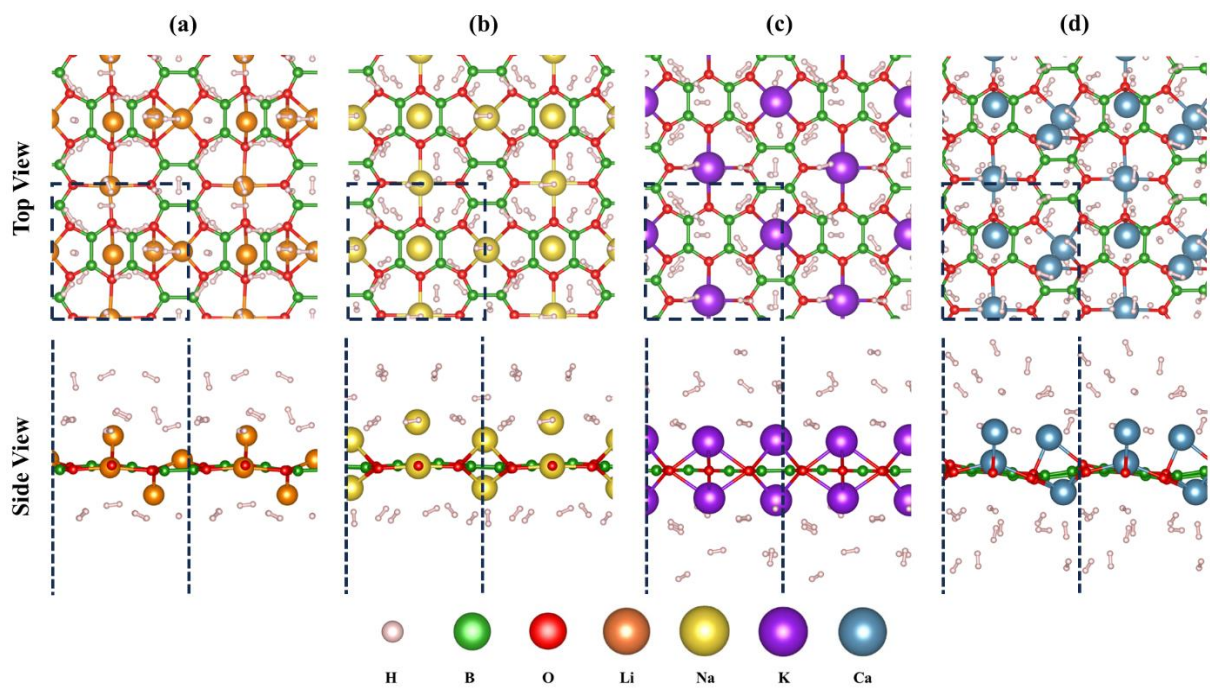

**Figure 6**

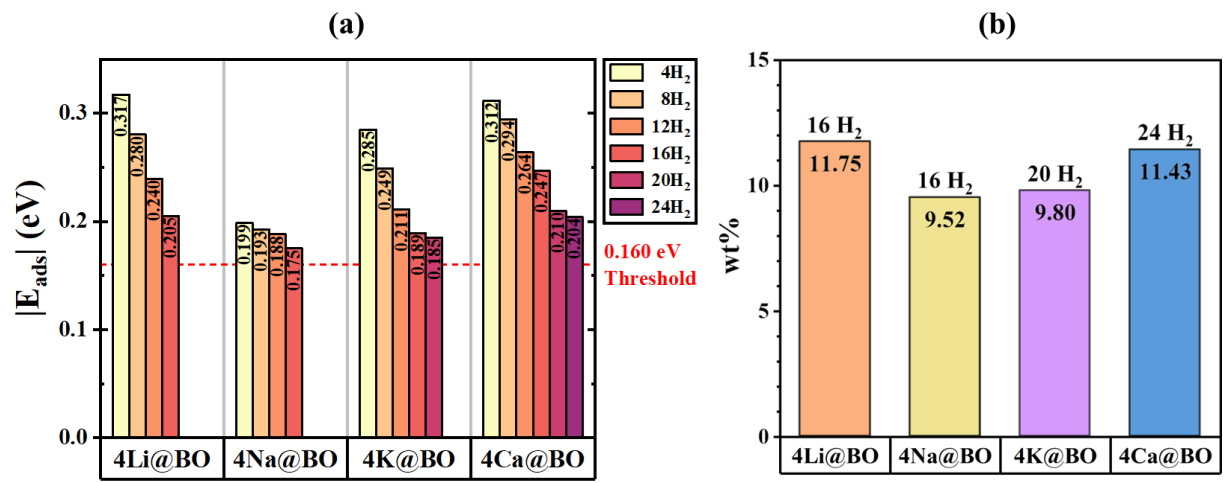

**Figure 7**

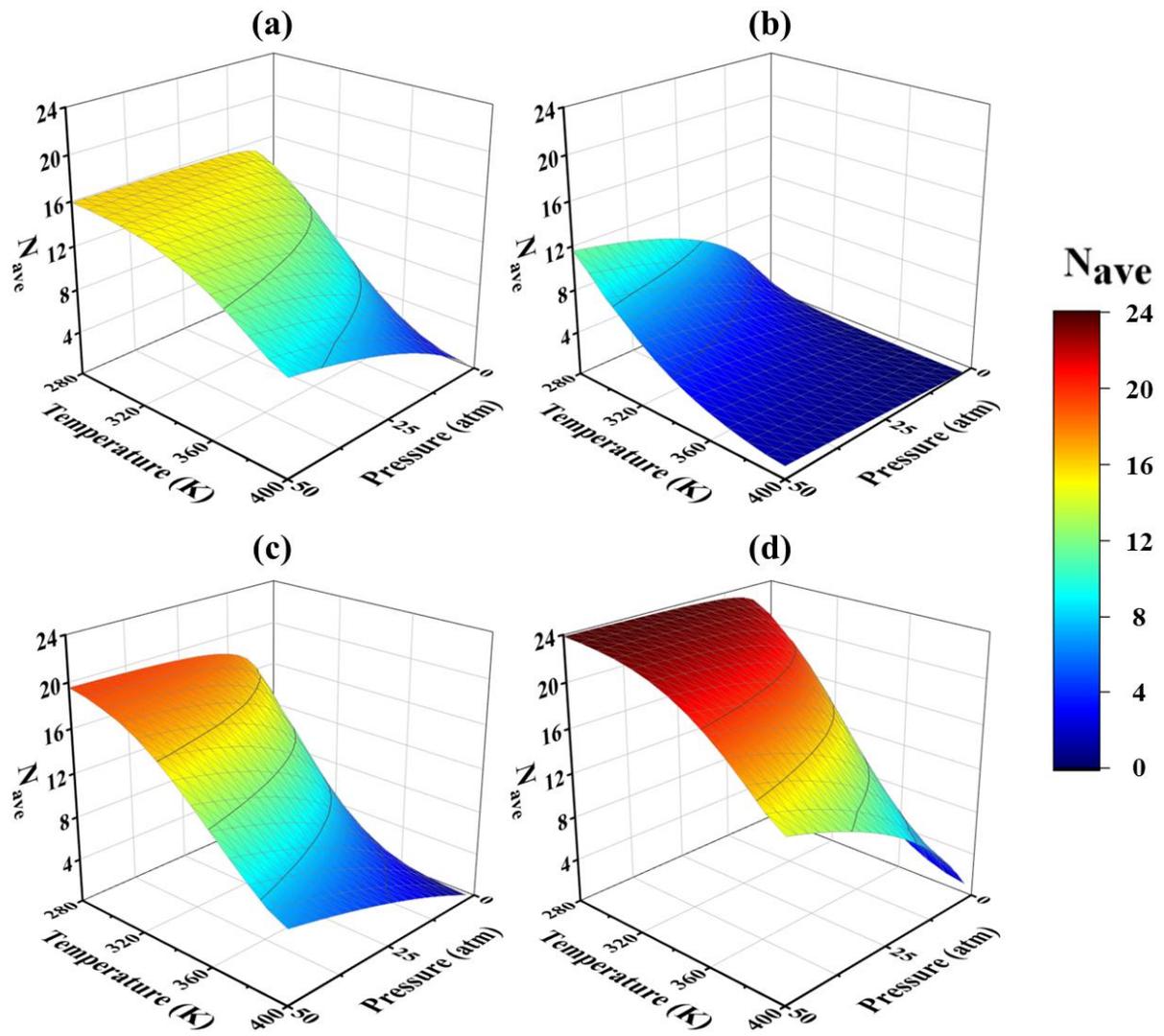

**Figure 8**